\documentclass{nature}

\usepackage{mathrsfs}
\usepackage{graphicx,graphics,color,epsfig}
\usepackage{bm}
\usepackage{amsmath}
\usepackage{amssymb}
\usepackage{subfigure}
\usepackage{amsfonts}
\usepackage{amstext}

\begin{document}

\title{Higgs-like Excitations of Cold Atom System with Spin-orbit Coupling}

\maketitle

\author{Fei-Jie Huang$^{\dagger}$,Qi-Hui Chen$^{\dagger}$,Wu-Ming Liu$^{\dagger\star}$}

\begin{affiliations}
\item
National Laboratory for Condensed Matter Physics,
Institute of Physics, Chinese Academy of Sciences,
Beijing 100190, China

$^\dagger$These authors contributed equally to this work.

$^\star$e-mail: wliu@iphy.ac.cn

\end{affiliations}

\begin{abstract}

The Higgs-like excitations, which distinguish from the Higgs amplitude mode
in many-body system, are single-particle excitations in system with non-Abelian gauge potential.
We investigate the Higgs-like excitations of cold atom system in artificial non-Abelian gauge potential.
We demonstrate that when a non-Abelian gauge potential is reduced to Abelian potential,
its Abelian part constructs spin-orbit coupling, and its non-Abelian part emerges
Higgs-like excitations. The Higgs-like excitations induce a mass of the
non-Abelian gauge field, which offsets the defect of massless of the gauge theories.
We show that the mass of gauge field can affect
the spin Hall currents which are produced by the spin-orbit coupling.
We also discuss the observation of these phenomena in real experiment.

\end{abstract}

The Higgs type excitations are significant for both particle physics and
condensed matter physics. In particle physics, the Higgs type excitations
generate the mass of particle. In condensed matter physics, the Higgs
type excitations relate to order phase of the system.
In particle physics, the suspected Higgs boson has been found by the Large
Hadron Collider (LHC) in 2012 \cite{Incandela2012, Gianotti2012}.
In condensed matter physics, the Higgs amplitude mode,
which is massive amplitude mode of collective excitations in many-body system, appears from
superconductors \cite{Lee2006} to superfluids
\cite{Volovik1992, Huber2008, Cooper2012, Pollet2012, Endres2012}.
However, all these results were concerned with the Higgs amplitude mode
in many-body systems, an investigation of the Higgs-like excitations is still absent.
The Higgs-like excitations are single-particle excitations in system
with non-Abelian gauge potential, the emergence of Higgs-like
excitations is due to the breaking of parallel transportation of non-Abelian gauge
potential, which connects to the spin-orbit (SO) coupling of the system.

The SO coupling, which is the interaction between the spin and the momentum of a
particle, is related to many effects in condensed matter physics
\cite{Murakami2003, Sinova2004, Hasan2010, Qi2011}.
Recently, with the realization of various artificial gauge potentials,
the cold atom system also can be designed to simulate
SO coupling \cite{Lin2011, Wang2012, Cheuk2012, Dalibard2011}.
This opens a new arena to explore novel
effects in cold atoms \cite{Ruseckas2005, Zhu2006, Liu2009, Wang2010, Iskin2011,
Bijl2011, Sinha2011, Hu2011, Liao2012, Deng2012, Ozawa2012, Zhang2012}.
The key of simulating SO coupling is the synthesis
of non-Abelian gauge potentials by engineering the
interaction between the atoms and the lasers \cite{Dalibard2011}.
It is well known that the non-Abelian gauge potentials
play a crucial role in understanding the fundamental interactions in particle physics.
In condensed matter physics, the non-Abelian gauge potentials also appear in study of
the mechanism of high-$T_{c}$ superconductivity \cite{Lee2006}
and graphene \cite{Vozmedianoa2010}.

In this report, we will investigate the Higgs-like excitations
of cold atom system in non-Abelian gauge potential. We first synthesize a $SU(2)$
gauge potential in cold atom system. By reducing the $SU(2)$
gauge potential to $U(1)$ potential, we analyze the
Abelian and non-Abelian parts of the gauge potential.
We find that the Abelian part of the gauge potential constructs
SO coupling, and its non-Abelian part emerges
Higgs-like excitations.
We show that the Higgs-like excitations give rise to
a mass of the gauge field, while the mass can affect the
spin Hall currents induced by SO coupling. Finally, we
discuss how to observe these phenomena through the detection
of spin Hall currents in real experiment.

\section*{Results}

\subsection{The spin-orbit coupling of cold atom system.}
We consider a cold atom system with each atom having a three-level $\Lambda$-type configuration
as shown in Fig. 1. Two ground states $\mid g_{1}\rangle$ and $\mid g_{2}\rangle$ are
coupled to an excited state $\mid e\rangle$ through the laser fields. The Rabi
frequencies are taken as $\Omega _{1}=\frac{1}{2}\Omega[\exp (i\mathbf{k}\cdot\mathbf{r})+\exp (i\mathbf{k'}\cdot\mathbf{r})]$
and $\Omega _{2}=\frac{1}{2i}\Omega[\exp (i\mathbf{k}\cdot\mathbf{r})-\exp (i\mathbf{k'}\cdot\mathbf{r})]$, in which $\mathbf{k}$
and $\mathbf{k'}$ are the wave vectors of lasers, $\mathbf{k'}=e^{i\varphi}\mathbf{k}$,
$\mathbf{r}$ is the position vector, $\varphi$ is an angle between the lasers as shown in Fig. 1b. The total
Rabi frequency is given by $\Omega =(\left\vert \Omega _{1}\right\vert
^{2}+\left\vert \Omega _{2}\right\vert ^{2})^{1/2}$. The Hamiltonian of cold atom
reads $H=H_{k}+H_{I}+V_{trap}$, in which $H_{k}=p^{2}/2m$ is the kinetic
energy, $p$ is the momentum, $m$ is the atomic mass, $V_{trap}$ is a trapping potential. The interacting
Hamiltonian is given by $H_{I}=2\Delta \mid e\rangle \langle e\mid +(\Omega
_{1}\mid e\rangle \langle g_{1}\mid +\Omega _{2}\mid e\rangle \langle g_{2}\mid
+h.c.)$, where $\Delta $ is the detuning.

For large detuning, by utilizing Berry phase, the eigenstates
of the interacting Hamiltonian will give rise to
an effective $SU(2)$ gauge potential (See Method section) $\mathbf{A}=\frac{1}{2}\mathbf{q}\sigma _{y}+\frac{1}{4}
\delta ^{2}\mathbf{Q}\sigma _{z}$, where $\sigma_{i}, i=x,y,z$ are the Pauli matrices,
$\mathbf{Q}=\mathbf{k}+\mathbf{k'}$ is the total wave vector of the laser fields
and $\mathbf{q}=\mathbf{k}-\mathbf{k'}$ is the relative wave vector,
$\delta =\arctan[(\frac{\Delta^{2}}{\Omega^{2}}+1)^{1/2}-\frac{\Delta}{\Omega}]$.
Redefining this effective potential to a $SU(2)$ scalar gauge potential $\mathcal{A}_{0}=$ $[
\mathbf{\gamma}, \mathbf{A}]_{+}$, where $[,]_{+}$ is
an anti-commutator, $\mathbf{\gamma} =\mathbf{p}/p$ is a dimensionless parameter.
Then neglecting the trapping potential and the constant terms,
the effective Hamiltonian of cold atom reads
\begin{equation}
H=H_{k}+g\mathcal{A}_{0}+V,
\end{equation}
in which $H_{k}=\eta p$ is the kinetic energy, $g=\eta $ is an effective
charge, $\eta =p/2m$ is a non-relativistic dimensionless
strength factor (assuming the light velocity $c=1$),
$V=\frac{1}{16m}\delta^{2}[q^{2}-(1+{\delta}^{2})
Q^{2}]\sigma _{z}$ $+\delta \frac{\Omega}{2}\sigma _{z}=M_{0}\sigma _{z}$ is
a scalar potential originating from construction of $\mathbf{A}$.

It is generally known that the SO coupling appears in $U(1)$ gauge potential, hence
to construct SO coupling, let's reduce the $SU(2)$ gauge potential $\mathcal{A}_{0}$ to
$U(1)$ potential. Firstly, we denote $
\mathbf{n}=(n_{x},n_{y},n_{z})$ as a unit vector which follows the direction of
magnetic field, and define the
direction vector of internal $SU(2)$ space as $\sigma(t)=n_{i}\sigma
_{i}(t)$, where $t$ is time, $i=x,y,z$. The definition implies that we have chosen the
same directions for the external and the internal spaces at the initial time. Then
along the direction vector $\sigma$, $\mathcal{A}_{0}$ will be
decomposed to two gauge potentials $\mathcal{A}
_{0}=\mathcal{A}+\mathcal{B}$ \cite{Faddeev1999, Niemi2006},
with
\begin{eqnarray}
&&\mathcal{A}=(\sigma\cdot \mathcal{A}_{0})\sigma+[\partial _{0}\sigma,\sigma],\nonumber\\
&&\mathcal{B}=[\sigma,\nabla _{0}\sigma],
\end{eqnarray}
where $\nabla _{0}=\partial _{0}+[\mathcal{A}_{0},]$ represents the time
component of covariant derivative, $[,]$ denotes the commutator. According to reduction
theorem, when $\mathcal{A}_{0}$ is reducible to $U(1)$ potential, there
should be
\begin{equation}
\nabla _{0}\sigma=0.
\end{equation}
Therefore, if $\mathcal{A}_{0}$
is a $U(1)$ potential, $\mathcal{A}_{0}=\mathcal{A}$. We view $\mathcal{A}$ is the Abelian part of
$\mathcal{A}_{0}$, and the potential $\mathcal{B}$ is the non-Abelian part.

Now let's discuss the physical meaning of the constraint condition
$\nabla _{0}\sigma=0$.
This condition actually is a parallel transportation.
Taking the Planck constant $\hbar =1$, $\partial _{0}=i
\partial _{t}$ will correspond to Hamiltonian operator.
Neglecting the scalar potential $V$ and
setting the effective charge $g=1$, we obtain $[\sigma,H]=[\sigma,\mathcal{A}_{0}]$, the
constraint condition becomes
\begin{equation}
i\partial _{t}\sigma=[\sigma,H].
\end{equation}
This equation is nothing but the equation of motion
of spin. Therefore, the constraint condition in fact corresponds to
the equation of motion
of spin, and the direction vector $\sigma$
corresponds to the spin operators of the system, as shown in Fig. 2.

We conclude that, only when the non-Abelian
gauge potential $\mathcal{A}_{0}$ satisfies the constraint condition$\
\nabla _{0}\sigma=0$, $\mathcal{A}_{0}$ is suited for constructing
SO coupling. If the constraint condition can not be satisfied, the gauge
potential $ \mathcal{B}$ in $\mathcal{A}_{0}$ will exist
and only $\mathcal{A}$ is suited for constructing SO
coupling in this case.

We now add the scalar potential $V$ to
the effective Hamiltonian to check whether $\mathcal{A}
_{0}$ satisfies the constraint condition
or not. The constraint condition can be
rewritten as $\nabla _{0}\sigma=[\sigma,H]+[
\mathcal{A}_{0},\sigma]$.
Because the scalar potential is a
non-Abelian potential, we find $\nabla _{0}\sigma\neq 0$.
In this situation, the SO coupling term of cold atom reads $H_{SO}=g
\mathcal{A}$, with its explicit expression
\begin{equation}
H_{SO}=\lambda \sigma _{y}+\nu \sigma _{z}.
\end{equation}
The coupling strength factors $\lambda $ and $\nu $ are $\lambda
=\frac{1}{m}[\frac{1}{2}\mathbf{q}\cdot\mathbf{p}+{\delta }^{3}\Omega p(\mathbf{Q}\cdot\mathbf{p}
\diagup \mathbf{q}\cdot\mathbf{p})]$ and $\nu =\frac{1}{m}(-2\delta \Omega p +2^{-2}\delta ^{2}\mathbf{Q}\cdot\mathbf{p})+\frac{1}{4m^{2}}\delta^{2}(Q^{2}-q^{2})p$,
respectively, in which the initial direction of $\sigma$
is pointed to $\mathcal{A}_{0}$ in the $SU(2)$ space.

\subsection{Higgs-like excitations of cold atom system.}
Since the non-Abelian gauge potential $\mathcal{A}_{0}$ does not satisfy
the constraint condition, there will be a residual term
left in the Hamiltonian after constructing SO coupling.
This term can be expressed as $H_{k,B}=g\mathcal{B}$,
with $\mathcal{B}=-\frac{\mathbf{Q}\cdot\mathbf{p}}{\mathbf{q}\cdot\mathbf{p}}
\delta ^{3}\Omega\sigma _{y}+[2\delta \Omega +\frac{1}{4m}\delta^{2}
(Q^{2}-q^{2})]\sigma _{z}$. The potential $\mathcal{B}$ is a
gauge covariant potential, it has a mass term $M^{2}_{B}Tr[\mathcal{B
}\cdot \mathcal{B}]$. Writing down the action of non-Abelian gauge field,
the mass reads $M_{B}=M_{0}/(1+(\mathbf{Q}\cdot\mathbf{p}
\diagup 2\mathbf{q}\cdot\mathbf{p})^{2}\delta ^{4})^{\frac{1}{2}}$ (See Method section).
Consequently, the effective Hamiltonian
of cold atom can be written as $H=H_{k}+H_{SO}+H_{B}$, in which
\begin{equation}
H_{B}=H_{k,B}+\eta^{\prime}M_{B}\sigma _{z}.
\end{equation}
It is clear that the Hamiltonian $H_{B}$ describes an excitation.
$H_{k,B}=g\mathcal{B}$ can be viewed as
the kinetic energy of the coupling between the excitation and the atom.
The scalar potential $V=\eta^{\prime }M_{B}\sigma _{z}$ in fact describes the coupling
between the mass of the excitation and the atom, where
$\eta^{\prime }=(1+(\mathbf{Q}\cdot\mathbf{p}
\diagup 2\mathbf{q}\cdot\mathbf{p})^{2}\delta ^{4})^{\frac{1}{2}}$ is a dimensionless
coupling factor.

Let's discuss the origin of the excitation. As shown in Fig. 2, the original direction of $
\sigma$ presents $SU(2)$ symmetry in the internal space, its
direction can be chosen freely. Nevertheless, when $\mathcal{A}
_{0}$ is used to construct SO coupling, the direction of $\sigma$
is confined to the constraint condition, and the initial symmetry is
broken. Yet in general, the whole $\mathcal{A}_{0}$
can not be suited for constructing SO coupling,
so there is a trend to restore the initial symmetry.
The potential $\mathcal{B}$ just plays the role of an
excited field which restores the initial symmetry.
We treat the excitation as Higgs-like excitation.
Due to the excitation, the gauge field $\mathcal{A}_{0}$ obtains mass.
Note that the excitations are different from the appearance of Higgs bosons
in dynamic gauge field. The appearance of Higgs bosons requires local gauge
invariance, while the gauge field discussed here is induced by Berry phase,
it is not dynamic and lack the local gauge invariance.

\subsection{The impact of Higgs-like excitations to spin Hall currents.}
The trajectories of single cold atom contributed by excitation can be calculated
from the equation of motion $\dot{\mathbf{r}}=-i[\mathbf{r},H_{B}]$.
As shown in Fig. 3a, the contribution includes two opposite trajectories.
The Fig. 3b shows the dispersion of $H_{B}$.
The dispersion is linear, and the energy does not vanish at zero momentum
due to the presence of excited mass $M_{B}$.

Next, let's discuss the impact of the excitations to the cold atom system.
A two-dimensional harmonic potential $\frac{1}{2}m\omega^{2}(y^{2}+z^{2})$ is chosen to trap the cold atoms.
The relationship between
the particle number $N$ and the trap frequency $\omega$ can be obtained by solving the equation
\begin{equation}
N=\int d\mathbf{r} n(r, t=0, T=0),
\end{equation}
in which
\begin{equation}
n(r, t, T)=\frac{1}{(2\pi)^{2}}
\int d\mathbf{p}[f_{+}(p, r, t, T)+f_{-}(p, r, t, T)]
\end{equation}
is the density profile of the cold atom system,
$f_{\sigma_{z}}=f_{\pm}(p, r, t, T)=[e^{\beta(H_{\pm}(p, r, t)-\mu)}+1]^{-1}$
are the spin-depending Fermi distributions with $\beta=1/k_{B}T$, $k_{B}$ and $T$ are the Boltzmann constant
and temperature. The Hamiltonian is given by (See Method section)
\begin{eqnarray}
H_{\pm}(p, r, t)=\frac{p^{2}}{2m}\pm(\frac{\mathbf{q}\cdot\mathbf{p}}{2m}+M_{B})+\frac{1}{2}m\omega^{2}z^{2}
+\frac{1}{2}m\omega^{2}(1\pm\frac{2m^{*}}{M_{B}})(y\pm\frac{\mathbf{q}\cdot\mathbf{p}}{2m}t)^{2}.
\end{eqnarray}
where $m^{*}=q^{2}/2m$ is a characteristic mass of the system.
The relationship is shown in Fig. 4.

We now discuss the spin Hall currents of the system.
In order to generate spin Hall currents, the wave vectors
of the lasers are chosen as $k_{x}=k'_{x}=0$,
the internal space is rotated $\pi/2$ around the $x$ direction. In this case,
the SO coupling reads $H_{SO}=\lambda\sigma _{z}$.
This term describes the spin Hall currents in which the
spin is polarized to $z$ direction while the currents move along $y$
direction. The spin Hall currents can be written as
\begin{equation}
J_{\sigma_{z}}^{y}=\frac{1}{(2\pi)^{2}}
\int d\mathbf{p}f_{\sigma_{z}}(p)j_{\sigma_{z}}^{y},
\end{equation}
where $j_{\sigma_{z}}^{y}=\langle
\hat{j}_{\sigma_{z}}^{y} \rangle$ are the single particle currents,
$\hat{j}_{\sigma_{z}}^{y}=\frac{1}{4}
[\sigma_{z},v_{y}]_{+}$ are the spin current operators, $v_{y}=-i[y,H]$ is the velocity along
$y$ direction. The impact of the
excitations to the spin Hall currents is shown in Fig. 5a.
It demonstrates that the spin down current is suppressed by the increase of the excited mass,
while the spin up current grows slightly.
The evolution of the atomic density profile $n(r,t)$ is shown in Fig. 5b.

\subsection{Experimental signatures of Higgs-like excitations.}
Let's discuss the observation of the Higgs-like excitations by detecting the spin Hall
currents. We can choose $^{6}Li$ atoms for a three-level $\Lambda$-type system,
the particle number is about $10^{4}$, a $2\pi\times10^{2}Hz$ harmonic
potential is used to trap the atoms.
The configuration of four laser fields is shown in Fig. 1b.
The wave number of the
lasers can be taken as $2\pi\times1.0(\mu m)^{-1}$ \cite {Cheuk2012}. For a large detuning, the Rabi
frequency and the detuning are required to satisfy $\Omega^{2}/\Delta\sim10^{6}Hz$.
When the laser fields are turned on, a non-Abelian gauge potential
is applied to the $^{6}Li$ atoms. By tuning the angle $\varphi$ between the lasers, different
masses of excitations can be obtained.

To detect the spin up current, we firstly initialize the atoms in $\mid g_{1}\rangle$ state.
Then a Raman pulse is applied between the states $\mid g_{1}\rangle$
and $\mid g_{2}\rangle$ to transfer the atoms to spin up state $\mid \chi _{D1}\rangle$.
The Rabi frequency of the pulse is required to match the spatial variation of
$\mid \chi _{D1}\rangle$. Turning on the lasers, the cold atoms will
experience the SO coupling, and couple to the excitations. After a time $t$, we
turn off the lasers and apply a reversal Raman pulse to transfer the atoms back
to the initial state. By using
the time of flight measurement,
the spin up current of the system can be determined. The measurement of the spin down current
(corresponding to the $\mid \chi _{D2}\rangle$ state)
also can be detected in the same manner \cite{Stanescu2007, Olson2013}.

\section*{Discussion}
In this report, we investigated the Higgs-like excitations,
which distinguish from the Higgs amplitude mode, in cold atom system.
In condensed matter physics, the Higgs amplitude
mode is collective excitations which are generated by interactions
between particles in many-body system,
the interactions induce the system to go
through a spontaneous breaking of continuous
symmetry, and the Higgs amplitude mode emerges.
Nevertheless, the Higgs-like excitations are single-particle excitations
which have nothing to do with interactions between particles,
as shown by Fig. 2, the emergence of Higgs-like
excitations is due to the breaking of parallel transportation of non-Abelian gauge
potential. Moreover, the Higgs amplitude mode
is independent of the SO coupling, while the Higgs-like excitations can
connect to SO coupling.

The impact of the Higgs-like
excitations to the spin Hall currents is shown in Fig. 5.
It demonstrates that the spin down current is suppressed,
while the spin up current grows slightly by the increase of the excited mass.
We discuss how to observe the Higgs-like
excitations through the detection of spin Hall currents in
real experiment.

It should be noted that the results we obtain not only confine in cold atom system,
but also can apply to systems concerning Berry phase.
We expect that the exploration of Higgs-like excitations
of cold atom system can help to understand the mass generation mechanism
both in the condensed matter physics and in particle physics.

\section*{Methods}
\noindent \textbf{$SU(2)$ gauge potential $\mathcal{A}_{0}$.}
The Hamiltonian of cold atom
reads $H=H_{k}+H_{I}+V_{trap}$, in which $H_{k}=p^{2}/2m$ is the kinetic
energy, $p$ is the momentum, $m$ is the atomic mass, $V_{trap}$ is a trapping potential. The interacting
Hamiltonian is given by $H_{I}=2\Delta \mid e\rangle \langle e\mid +(\Omega
_{1}\mid e\rangle \langle g_{1}\mid +\Omega _{2}\mid e\rangle \langle g_{2}\mid
+h.c.)$, where $\Delta $ is the detuning.

Defining $\mathbf{Q}=\mathbf{k}+\mathbf{k'}$ as the total wave vector of the laser fields,
$\mathbf{q}=\mathbf{k}-\mathbf{k'}$ as the relative wave vector,
$\delta =\arctan[(\frac{\Delta^{2}}{\Omega^{2}}+1)^{1/2}-\frac{\Delta}{\Omega}]$,
$\phi=\frac{1}{2}\mathbf{Q}\cdot\mathbf{r}$, $\theta=\frac{1}{2}\mathbf{q}\cdot\mathbf{r}$,
then the eigenvectors of interacting Hamiltonian can be expressed as
\begin{eqnarray}
&&|\chi _{D1}\rangle= -\sin\theta|g_1\rangle + \cos\theta|g_2\rangle,\nonumber\\
&&|\chi _{D2}\rangle= -\exp(i\phi)\sin\delta|e\rangle + \cos\delta \cos\theta|g_1\rangle + \cos\delta \sin\theta|g_2\rangle,\nonumber\\
&&|\chi _3\rangle = \exp(i\phi)\cos\delta|e\rangle + \sin\delta \cos\theta|g_1\rangle + \sin\delta \sin\theta|g_2\rangle,
\end{eqnarray}
with the eigenvalues are $0$, $\frac{\Omega}{2\Delta}$, $2\Delta$, respectively.

For large detuning, we can neglect the eigenvalue $2\Delta$ and assume the other two eigenvalues
are degeneracy, hence by utilizing Berry phase, the remaining two eigenvectors
of the interacting Hamiltonian will give rise to
an effective $SU(2)$ gauge potential $\mathbf{A}_{\alpha\beta}=-i\langle\chi _{\alpha}|\nabla|\chi _{\beta}\rangle$,
$\alpha, \beta=D1, D2$,
with its explicit expression
\begin{equation}
\mathbf{A}=\frac{1}{2}\mathbf{q}\sigma _{y}+\frac{1}{4}
\delta ^{2}\mathbf{Q}\sigma _{z},
\end{equation}
where $\sigma_{i}, i=x,y,z$ are the Pauli matrices.
Redefining potential $\mathbf{A}$, we obtain $SU(2)$ gauge potential
\begin{equation}
\mathcal{A}_{0}=[\mathbf{\gamma}, \mathbf{A}]_{+},
\end{equation}
where $[,]_{+}$ is
an anti-commutator, $\mathbf{\gamma} =\mathbf{p}/p$ is a dimensionless parameter.

\noindent \textbf{The mass of the gauge field.}
Because the gauge field has only time component
$\mathcal{A}_{0}$, we can expand the action of the gauge field as
\begin{equation}
S=TrF_{0i}F_{0i}+\frac{1}{2}TrF_{00}F_{00},
\end{equation}
where $0$ is the time component index and $i$ are space component indices,
respectively. It is obvious that the second term of the expansion is the
mass term. When $\mathcal{A}_{0}$ is substituted to the mass term, we have
\begin{equation}
\frac{1}{2}TrF_{00}F_{00}=M^{2}_{B}Tr[\mathcal{B}\cdot \mathcal{B}],
\end{equation}
in which the mass reads
\begin{equation}
M_{B}=M_{0}/(1+(\mathbf{Q}\cdot\mathbf{p}
\diagup 2\mathbf{q}\cdot\mathbf{p})^{2}\delta ^{4})^{\frac{1}{2}}.
\end{equation}

\noindent \textbf{Derivation of the Hamiltonian (9).}
The wave vectors of the lasers are chosen as $k_{x}=k'_{x}=0$,
the internal space is rotated $\pi/2$ around the $x$ direction. Combining the Eq. (5) and Eq. (6), the explicit
expression of Hamiltonian can be written as
\begin{equation}
H(p, r)=\frac{p^{2}}{2m}+\frac{\mathbf{Q}\cdot\mathbf{p}}{4m}\delta ^{2}\sigma _{y}
+\frac{\mathbf{q}\cdot\mathbf{p}}{2m}\sigma _{z}+\eta^{\prime}M_{B}\sigma _{z}
+\frac{1}{2}m\omega^{2}(y^{2}+z^{2}).
\end{equation}
Diagonalizing the Hamiltonian and expanding in $M_{B}$ order, we have
\begin{equation}
H_{\pm}(p, r)=\frac{p^{2}}{2m}\pm(\frac{\mathbf{q}\cdot\mathbf{p}}{2m}+M_{B})
+\frac{1}{2}m\omega^{2}z^{2}+\frac{1}{2}m\omega^{2}(1\pm\frac{2m^{*}}{M_{B}})y^{2},
\end{equation}
in which $m^{*}=q^{2}/2m$ is a characteristic mass of the system.
From $\dot{\mathbf{r}}=-i[\mathbf{r},H]$, the time evolved Hamiltonian can be obtained
\begin{equation}
H_{\pm}(p, r, t)=\frac{p^{2}}{2m}\pm(\frac{\mathbf{q}\cdot\mathbf{p}}{2m}+M_{B})+\frac{1}{2}m\omega^{2}z^{2}
+\frac{1}{2}m\omega^{2}(1\pm\frac{2m^{*}}{M_{B}})(y\pm\frac{\mathbf{q}\cdot\mathbf{p}}{2m}t)^{2}.\tag{9}
\end{equation}



\begin{addendum}

\item [Acknowledgement]

This work was supported by NKBRSFC
under grants Nos. 2011CB921502, 2012CB821305,
the NSFC under grants Nos. 61227902.

\item [Author Contributions]
All authors planned and designed theoretical numerical studies.
All contributed in completing the paper.

\item [Competing Interests]
The authors declare that they have no competing financial interests.

\item [Correspondence]
Correspondence and requests for materials should be addressed to Liu, Wu-Ming.
\end{addendum}

\clearpage

\newpage

\bigskip

\textbf{Figure 1 The configuration of three-level $\Lambda $-type atoms interacting with laser fields.}
(a) Two ground states $\mid g_{1}\rangle$ and $\mid g_{2}\rangle$ are
coupled to an excited state $\mid e\rangle$ through the laser fields.
$\Omega _{1}$ and $\Omega _{2}$ are the Rabi frequencies, $\Delta$ is a
large detuning. (b) Two inner laser fields (the blue arrows)
are arranged to form Rabi frequency $\Omega _{1}$,
other two lasers (the red arrows) are arranged to form $\Omega _{2}$. There is
an angle $\varphi$ between the laser fields.

\bigskip

\textbf{Figure 2 The sphere surface of $SU(2)$ gauge potential $\mathcal{A}_{0}$.}
(a) At the initial time $t=0$, the basic vectors of the gauge potential $\sigma_{x}(0)$, $\sigma_{y}(0)$ and $\sigma_{z}(0)$ point to certain directions. The direction vector (the red arrow) $\sigma(0)$ points to $A$. (b) After a time $t$, the basic vectors change to $\sigma_{x}(t)$, $\sigma_{y}(t)$ and $\sigma_{z}(t)$ directions.
The direction vector $\sigma(t)$ changes along with the basic vectors and points to $B$. If the path $AB$ satisfies the equation of motion of spin, the SO coupling will present. If not, then excepting for SO coupling, Higgs-like excitation will emerge.

\bigskip

\textbf{Figure 3 The impact of Higgs-like excitation to single cold atom.}
(a) The trajectories of a single cold atom contributed by Higgs-like excitation.
The red solid line and the blue dash line correspond to the velocities
$v=3\times10^{-3}\mu m/s$ and $v=5\times10^{-3}\mu m/s$, respectively.
(b) The dispersion of coupling energy between the Higgs-like excitation and a single atom, $\varepsilon_{B}=(\epsilon p+\eta^{\prime }M_{B})\sigma _{z}$.
The red solid line and the blue dash line correspond to the coupling strength parameters
$\epsilon=4m^{-1}\times10^{-2}$ and $\epsilon=6m^{-1}\times10^{-2}$, respectively.
The unit of the energy $\varepsilon_{B}$ is the recoil energy ${\varepsilon}_{R}=k^{2}/(2m)$.

\bigskip

\textbf{Figure 4 The relationship between the particle number $N$ and the trap frequency $\omega$.}
$\varepsilon_{F}$ is the Fermi energy. The red solid line and the blue dash line correspond
to the ratios $M_{B}/m^{*}=2.2$ and $M_{B}/m^{*}=2.8$, respectively.

\bigskip

\textbf{Figure 5 The impact of Higgs-like excitations to spin Hall currents.}
(a) The relationship between the
excited mass $M_{B}$ and the spin Hall currents $J_{\sigma_{z}}^{y}$, in which
the spin is polarized to $z$ direction while the currents move along $y$
direction. The red (blue) line corresponds to the spin up (down) current.
The unit of $J_{\sigma_{z}}^{y}$ is $\frac{1}{2\pi}\times10 cm/s$.
(b) The evolution of the atomic density profile from the time $t=0ms$ to $t=4ms$ and $t=9ms$.
The red and blue parts in each figure denote the spin up and spin down atoms.
The unit of $y$ and $z$ directions is $\mu m$.
The temperature is taken as $T=0.4T_{F}$, $T_{F}$ is the Fermi temperature.
The up and down figures correspond to the ratios
$M_{B}/m^{*}=2.2$ and $M_{B}/m^{*}=2.8$, respectively.
The figures show the trends of spin currents of Fig. 5a.

\newpage

\begin{figure}
\begin{center}
\epsfig{file=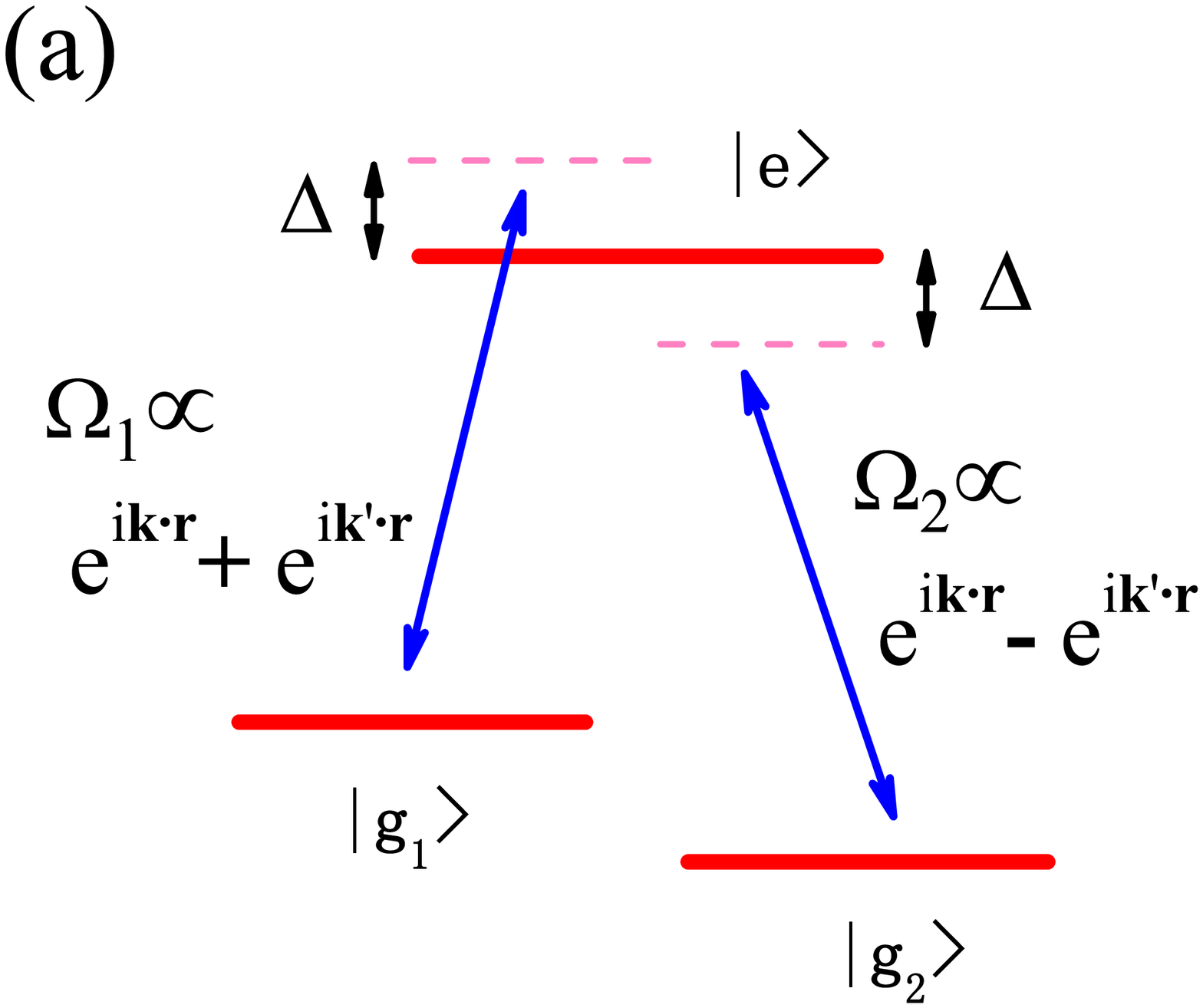,width=12cm}
\epsfig{file=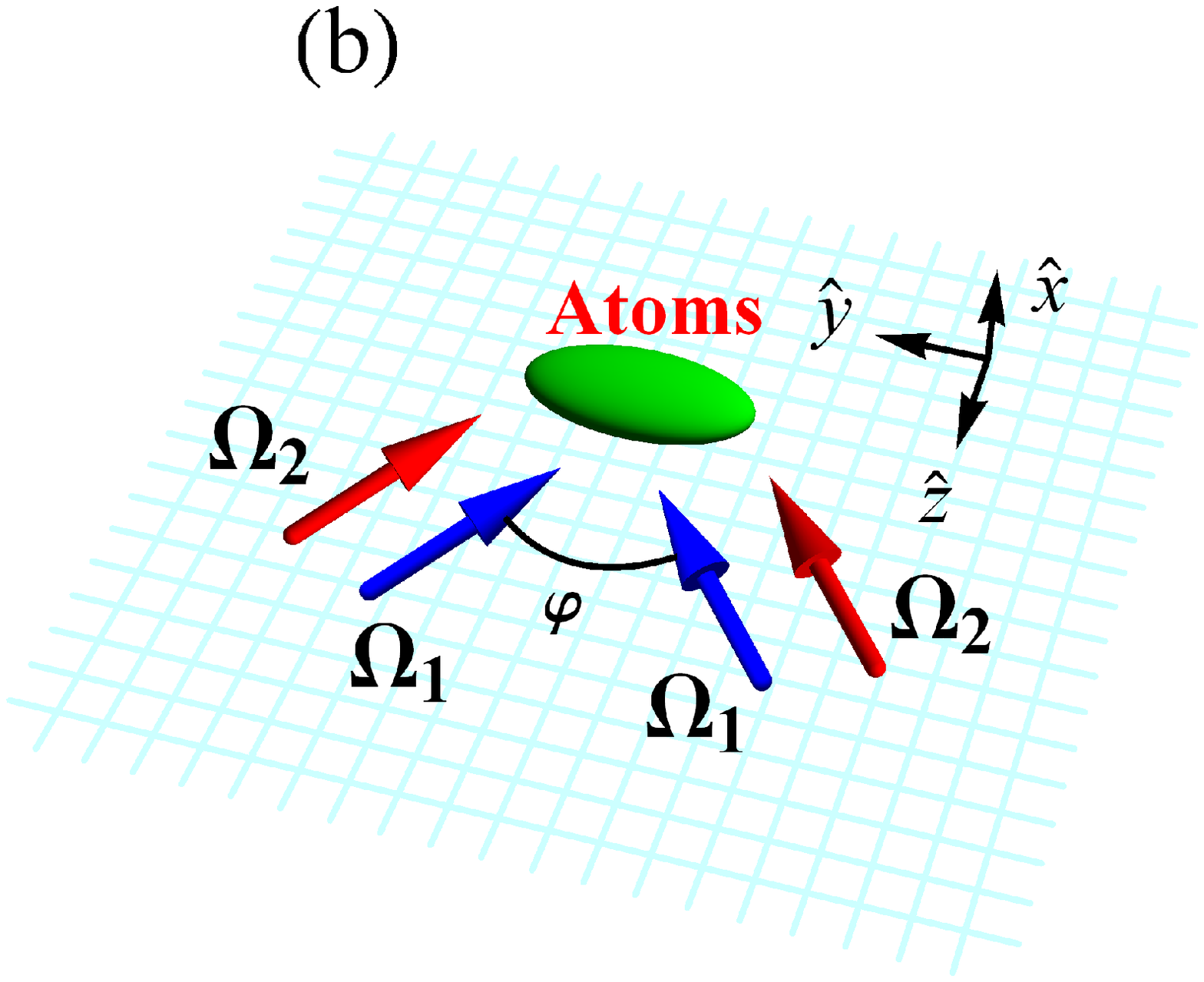,width=10cm}
\end{center}
\label{Fig. 1}
\end{figure}

\begin{figure}
\begin{center}
\epsfig{file=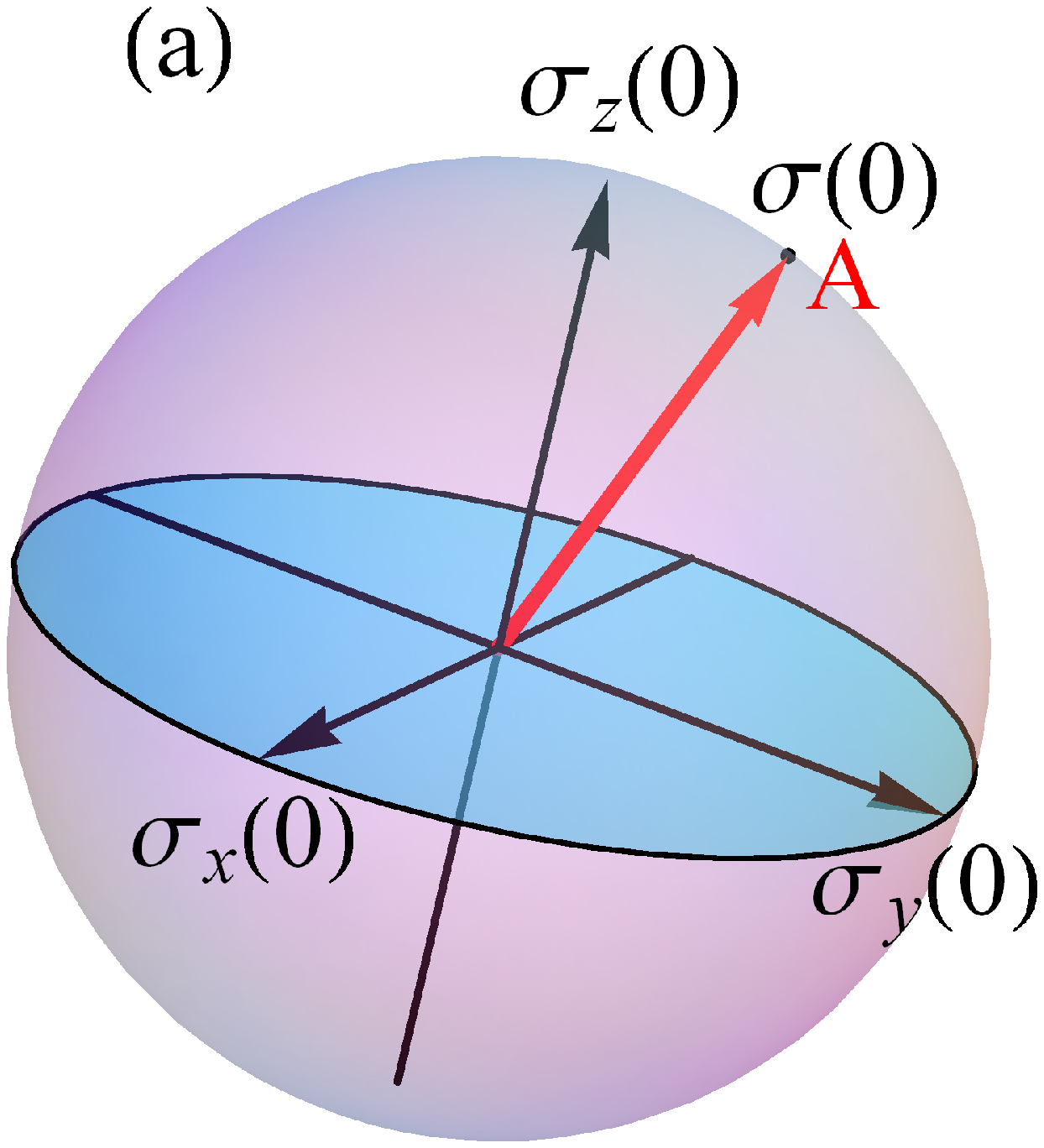,width=10cm}
\epsfig{file=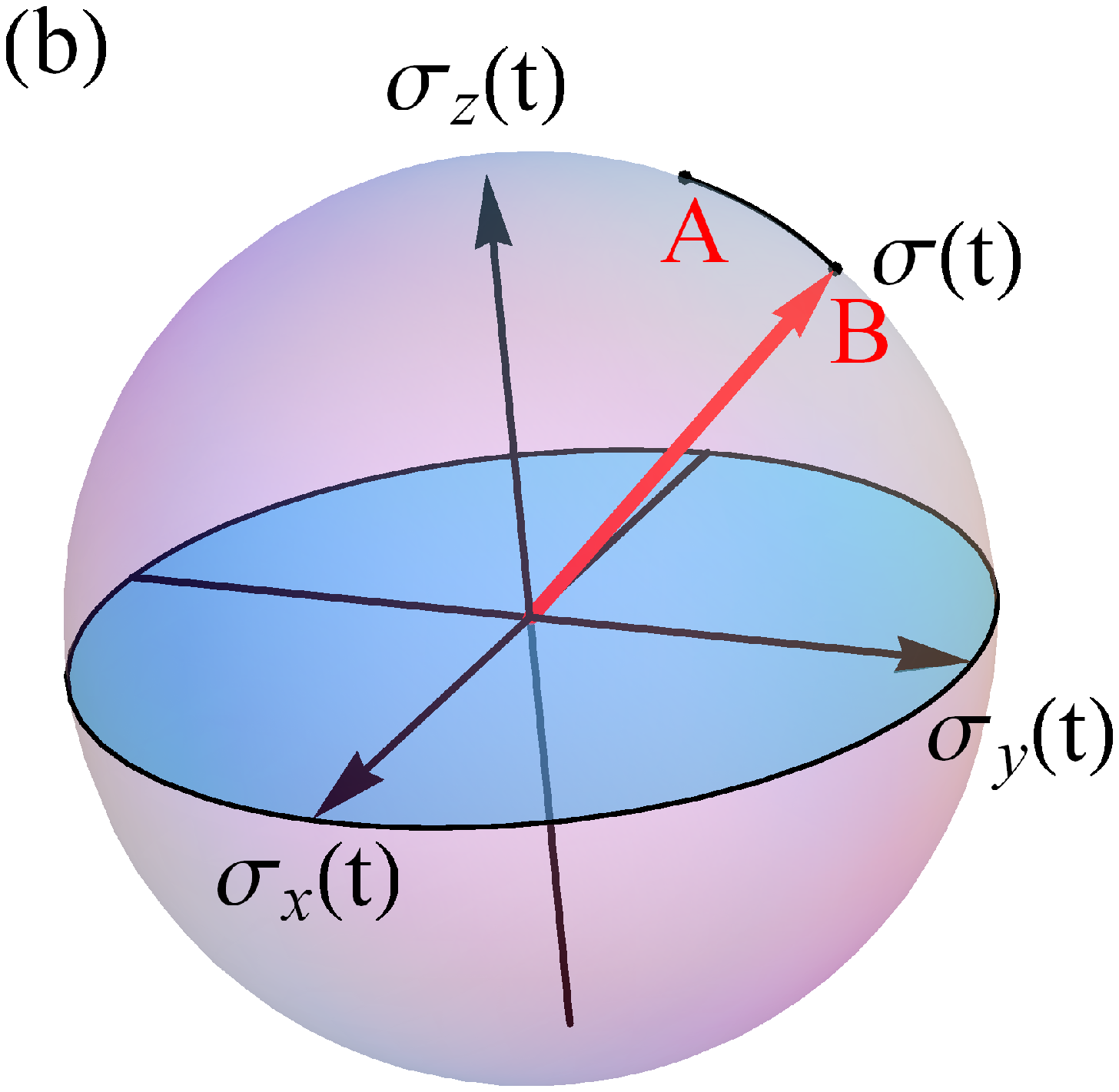,width=10cm}
\end{center}
\label{Fig. 2}
\end{figure}

\begin{figure}
\begin{center}
\epsfig{file=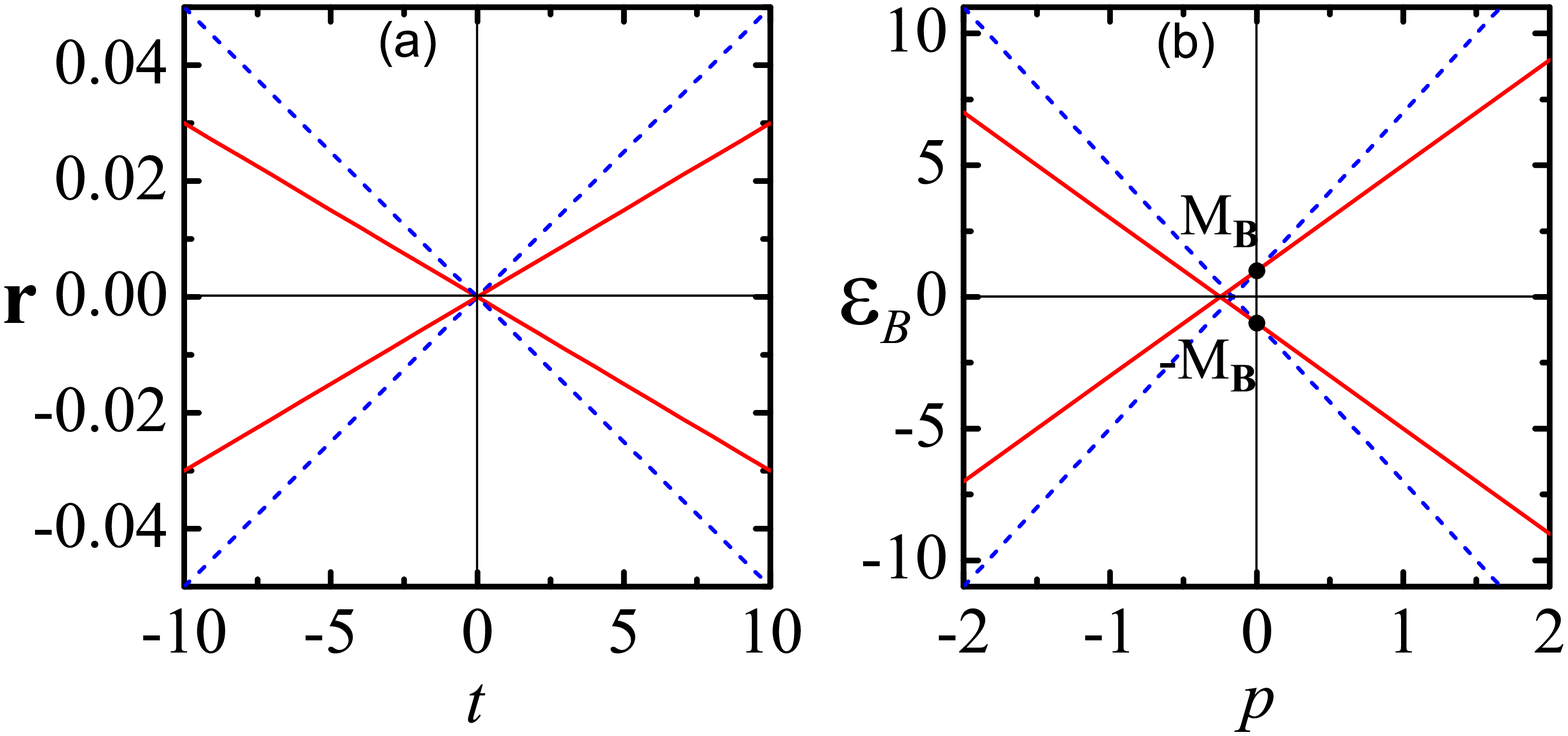,width=14cm}
\end{center}
\label{Fig. 3}
\end{figure}

\begin{figure}
\begin{center}
\epsfig{file=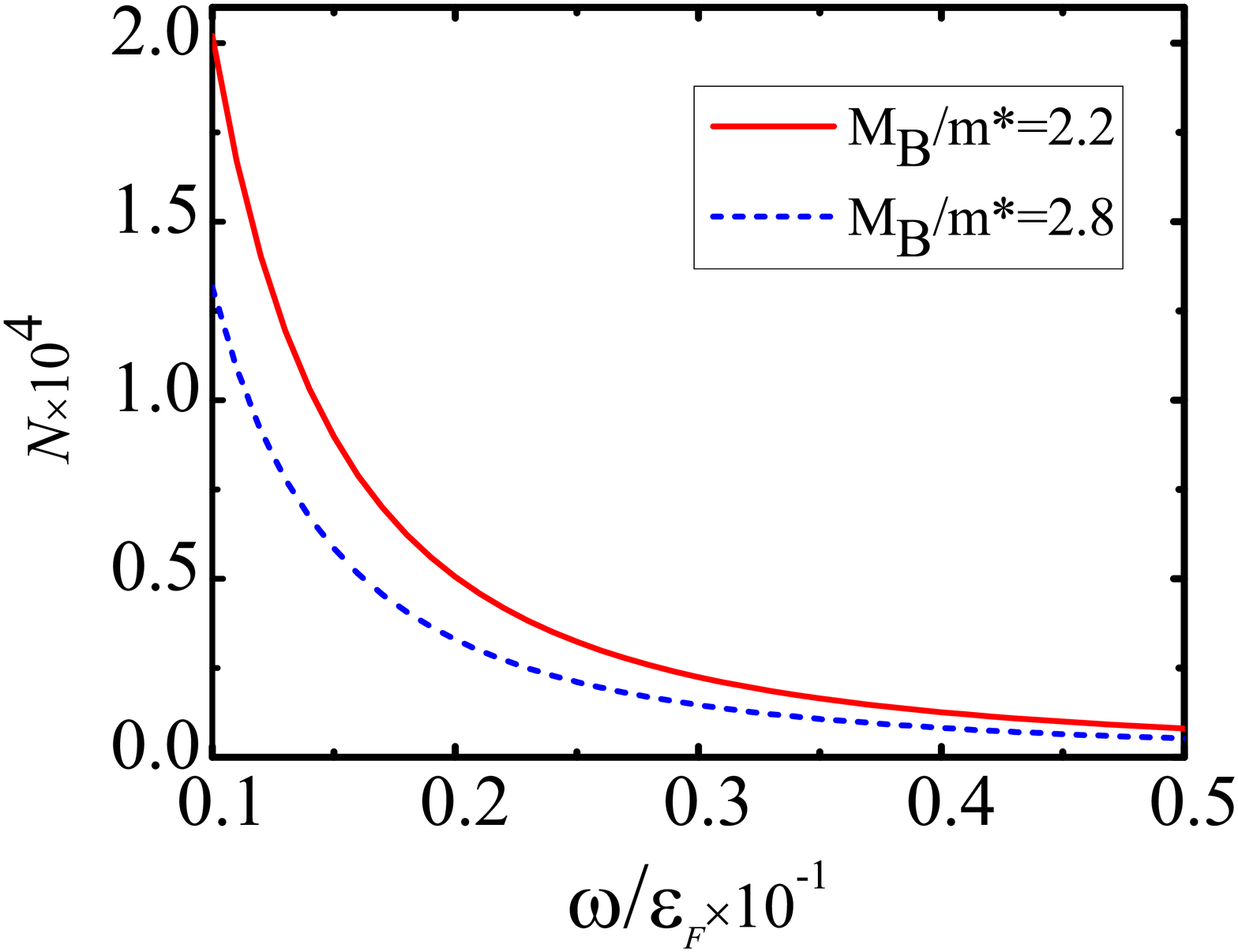,width=14cm}
\end{center}
\label{Fig. 4}
\end{figure}

\begin{figure}
\begin{center}
\epsfig{file=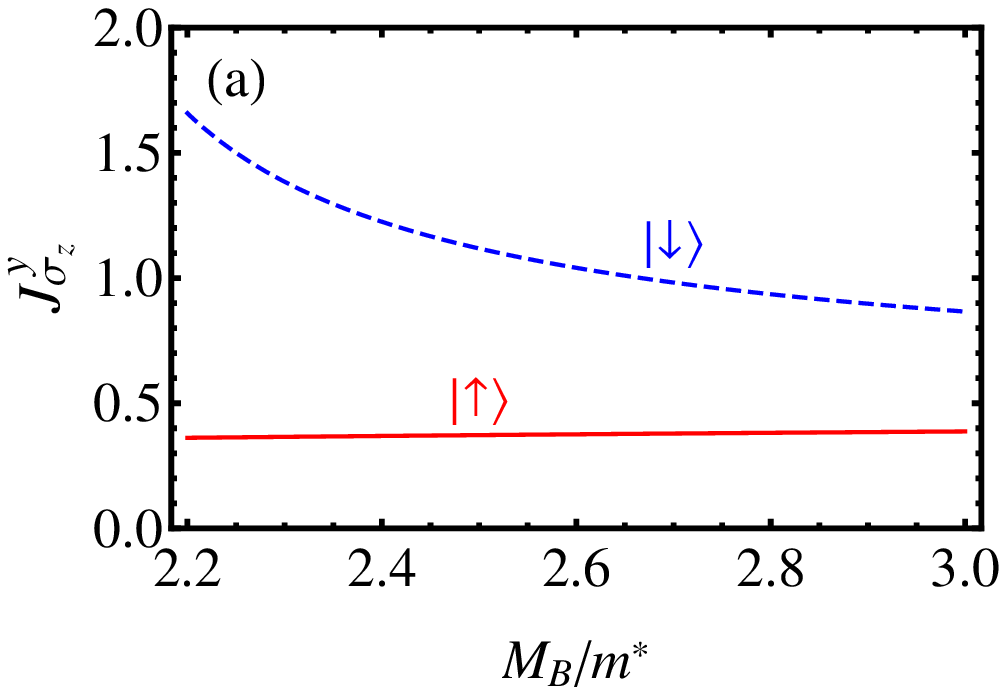,width=10.2cm}
\epsfig{file=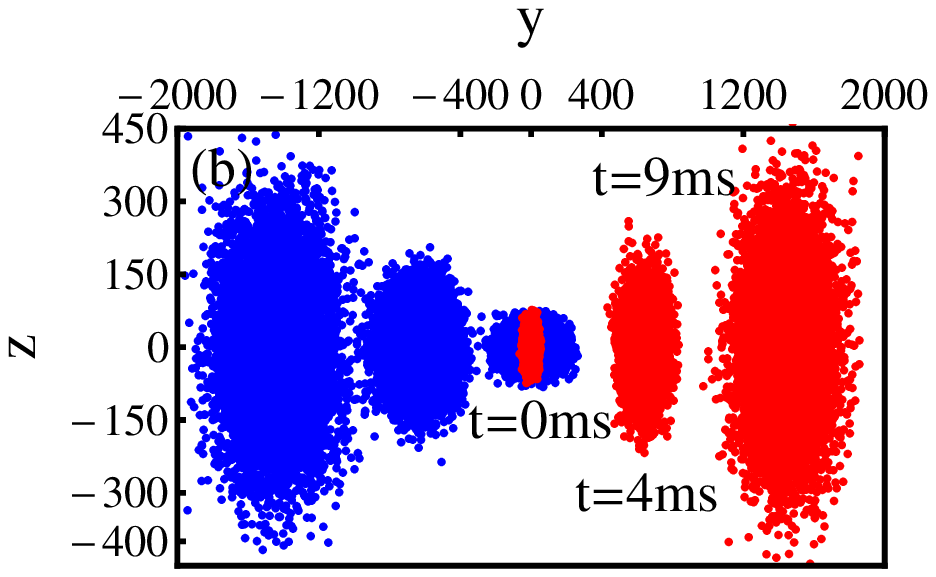,width=12cm}
\epsfig{file=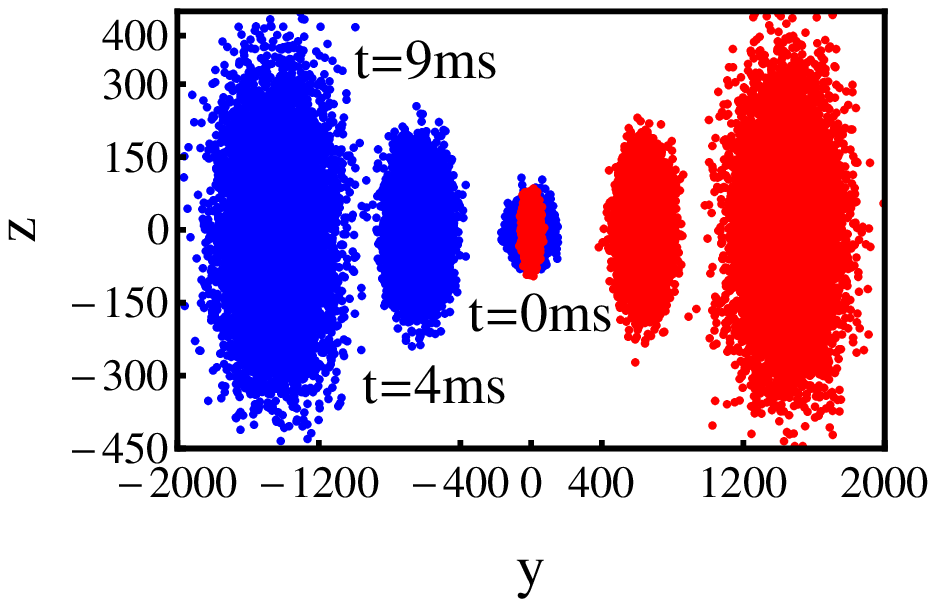,width=12cm,trim=0in 0in 0in 0.75in}
\end{center}
\label{Fig. 5}
\end{figure}

\end{document}